\documentclass{vgtc}
\usepackage{amsmath,amsfonts}
\usepackage{algorithmic}
\usepackage{algorithm}
\usepackage{booktabs}
\usepackage{array}
\usepackage{textcomp}
\usepackage{stfloats}
\usepackage{url}
\usepackage{verbatim}
\usepackage{graphicx}
\usepackage{cite}
\usepackage{comment}
\usepackage{soul}
\usepackage{transparent}
\usepackage[table,svgnames]{xcolor}
\usepackage{tcolorbox}
\usepackage{setspace}
\usepackage{tikz}
\usepackage{amssymb}
\usepackage{fancyhdr}
\usetikzlibrary{calc,positioning,arrows.meta}
\usepackage{wrapfig}
\usepackage{endnotes}
\usepackage{minted}
\usepackage{multirow}
\usepackage{xcolor, colortbl}
\usepackage{enumitem}
\definecolor{OliveGreen}{RGB}{183, 207, 178}
\definecolor{DarkYellow}{HTML}{DEA601}
\definecolor{DarkOrange}{HTML}{ED7D31}
\definecolor{DarkBlue}{HTML}{4472C4}
\usepackage{xspace}

\definecolor{tsblue}{HTML}{2C3E50}   % for \nonterm
\definecolor{tsteal}{HTML}{2F855A}   % for \literal
\definecolor{tsorange}{HTML}{C05621} % for \typeval

\usepackage[compact]{titlesec}
\titlespacing{\section}{0pt}{1ex}{0ex}
\titlespacing{\subsection}{0pt}{1ex}{0ex}

\makeatletter
\renewcommand\section{\@startsection{section}{1}{\z@}%
  {-2ex \@plus -1ex \@minus -.2ex}%
  {0.8ex \@plus .2ex}%
  {\@afterindentfalse\reset@font\normalsize\sffamily\bfseries\scshape\vgtc@sectionfont}}

\renewcommand\subsection{\@startsection{subsection}{2}{\z@}%
  {-1.8ex\@plus -1ex \@minus -.2ex}%
  {0.8ex \@plus .2ex}%
  {\@afterindentfalse\reset@font\normalsize\sffamily\bfseries\vgtc@sectionfont}}

\renewcommand\subsubsection{\@startsection{subsubsection}{3}{\z@}%
  {-1.8ex\@plus -1ex \@minus -.2ex}%
  {0.8ex \@plus .2ex}%
  {\@afterindentfalse\reset@font\sffamily\normalsize\vgtc@sectionfont}}
\makeatother

\onlineid{0}

\vgtccategory{Application}

\vgtcpapertype{application}

\newcommand{\papertitle}{A Case Study in Accessible Redesign of a Wastewater Dashboard}
\title{\papertitle}

\author{%
  \authororcid{Tingying He}{0000-0002-9670-5587},
  \authororcid{Jake Wagoner}{0009-0000-5053-2281},
  \authororcid{Md Rahat-uz- Zaman}{0000-0001-6728-7569},
  \authororcid{Md Dilshadur Rahman}{0009-0008-5467-615X},
  \authororcid{Willy Ray}{},\\
  \authororcid{George G. Vega Yon}{0000-0002-3171-0844},
  \authororcid{Matthew Samore}{0000-0002-4862-9196},
  \authororcid{Alexander Lex}{0000-0001-6930-5468},
  \authororcid{Paul Rosen}{0000-0002-0873-9518}
}

\authorfooter{
    \item T. He, J. Wagoner, M. Rahat-uz-Zaman, M. D. Rahman, W. Ray, G. G. Vega Yon, M. Samore, and P. Rosen are with the University of Utah. A. Lex is with Graz University of Technology and the University of Utah.
    \item T. He's E-mail: tingying.x.he@gmail.com
}

\abstract{
  Public health dashboards communicate data that can inform important decisions, but they often raise accessibility challenges. We present a case study redesigning the Utah Wastewater Surveillance System dashboard to improve accessibility and usability across desktop and mobile settings. The redesign was informed by WCAG and developed iteratively with Utah DHHS collaborators, and we collected feedback on the final design from an external blind researcher. Our case study highlights that accessible dashboard design requires aligning accessibility guidelines with user needs, stakeholder workflows, and technical constraints. It also suggests that simple, targeted technical solutions tailored to the existing environment can provide practical value for dashboard redesign in government contexts.
}


\graphicspath{{figs/}{figures/}{pictures/}{images/}{./}} % where to search for the images

\usepackage{tabu}                      % only used for the table example
\usepackage{booktabs}                  % only used for the table example
\usepackage{lipsum}                    % used to generate placeholder text
\usepackage{mwe}                       % used to generate placeholder figures
\usepackage{ccicons}                   % package to be able to use icons from creative commons

\usepackage{mathptmx}                  % use matching math font

\usepackage{tabularx}
\usepackage{multirow}

\usepackage{cuted} % appendix
\usepackage{caption} % for \captionof

\newcommand{\rev}[1]{\textcolor{SeaGreen}{#1}} %reivision for VIS26
\renewcommand{\rev}[1]{\textcolor{Black}{#1}} %reivision for VIS26

\begin{document}

%%%%%%%%%%%%%%%%%%%%%%%%%%%%%%%%%%%%%%%%%%%%%%%%%%%%%%%%%%%%%%%%
%%%%%%%%%%%%%%%%%%%%%% START OF THE PAPER %%%%%%%%%%%%%%%%%%%%%%
%%%%%%%%%%%%%%%%%%%%%%%%%%%%%%%%%%%%%%%%%%%%%%%%%%%%%%%%%%%%%%%%

% \firstsection{Introduction}

\maketitle

\setstretch{0.96}
%% \section{Introduction} %for journal use above \firstsection{..} inste

\section{Introduction} %for journal use above \firstsection{..} instead

Dashboards are a common use case for visualizations, which typically feature interactive interfaces and multiple data views~\cite{Sarikaya:2019:What}. 
In public health, they are widely used to monitor changing conditions and communicate surveillance data to policymakers, domain experts, and the public. Because these dashboards can inform important decisions, they should be accessible to all users.
This need has become more urgent in the USA as accessibility requirements for public digital services have become more specific.
In 2024, updates to Title II of the Americans with Disabilities Act (ADA) established specific accessibility requirements for web content and mobile applications provided by state and local governments~\cite{doj:2024:fact}.

In this paper, we report on a case study redesigning the Utah Department of Health and Human Services (DHHS) Wastewater Surveillance system \rev{to support current accessibility requirements and improve access for disabled individuals}, particularly those with color vision deficiencies, and those who are blind or have low vision (BLV).
In addition, we considered accessibility not only for disabled users but for all users, so we also emphasized improving website usability and making visualizations understandable to the general public.

Our redesign process combined accessibility review, visualization design, stakeholder collaboration, and implementation under practical government constraints. 
We reviewed the existing dashboard, identified accessibility and usability barriers, proposed alternative designs, and iteratively refined the solutions with DHHS collaborators. We further extended this redesign to improve accessibility and usability on mobile devices. We evaluated our redesign with our collaborators and collected feedback from a blind researcher.

\section{Related Work}

Dashboard design is a key research area in visualization. Prior work has developed guidelines, design spaces, and heuristics to support dashboard creation (e.g., \cite{Sarikaya:2019:What, Bach:2023:Dashboard, Setlur:2024:Heuristics}).
Researchers have also explored computational support for dashboard authoring and recommendations, such as MEDLEY~\cite{Pandey:2023:MEDLEY}  and DMiner~\cite{Lin:2024:DMiner}. 

Dashboards are widely used in public health and have attracted researchers' attention~\cite{Schulze:2023:Digital,Stahlman:2025:Design}. 
Prior work proposed design guidelines for public health dashboards (e.g., \cite{Rabiei:2024:Developing, Dasgupta:2022:Future, Lechner:2014:Towards, Ansari:2022:Development}) and investigated the generation of public health dashboards (e.g., QualDash~\cite{Elshehaly:2021:QualDash}).
During the COVID-19 pandemic, dashboards became especially prominent, which prompted substantial research on their design and use for tracking and communicating COVID-19 data (e.g., \cite{Dong:2020:Interactive, Zhang:2023:Visualization, Arleo:2025:Reflections, Clarkson:2023:Webs}).

Accessibility remains a key challenge for dashboards. For example, because conventional visualizations rely heavily on the visual system, their content can be difficult for blind and low-vision users to access. Researchers have identified the challenges of the accessibility of visualization~\cite{Kim:2021:Accessible} and provided evaluation guidelines~\cite{Elavsky:2022:How}. The community has developed a range of approaches to support BLV users' access to visualization content, including alternative text~\cite{Zong:2022:Rich, McNutt:2025:Accessible, Lundgard:2022:Accessible, Smits:2024:AltGosling}, tactile and haptic representations~\cite{Yang:2020:Tactile, Engel:2019:User, Goncu:2011:GraVVITAS, He:2025:InTouch, He:2026:Using}, sonification~\cite{Hoque:2023:Accessible, Daunys:2008:Sonification, Franklin:2003:Pie}, or interactive chart exploration support~\cite{Elavsky:2024:Data, Blanco:2022:Olli, Gorniak:2024:VizAbility, Seo:2024:MAIDRAI, Seo:2024:MAIDR,He:2027:Touching}.

Visualization accessibility challenges are amplified in dashboards, which commonly combine multiple views and interactive controls.
Prior work has identified gaps for supporting screen-reader users in using interactive visualizations~\cite{Kim:2021:Accessible} and navigating across multiple views~\cite{Siu:2021:Covid, Fan:2023:Accessibility}.
Recent research has begun to address these issues specifically in dashboard contexts. 
For example, Srinivasan et al.\ used a co-design method with screen-reader users to identify accessibility challenges and design goals, and developed Azimuth, a prototype for optimizing dashboards for accessibility \cite{Srinivasan:2023:Azimuth}.

\section{Project Background}

The Utah Wastewater Surveillance System\footnote{Utah Wastewater Surveillance system (SARS-CoV-2 tab): \href{https://dhhs.utah.gov/health-dashboards/utah-wastewater-surveillance-system/}{dhhs.utah.gov/health-dashboards/utah-wastewater-surveillance-system/}.} is a public-facing health dashboard managed by DHHS and part of its broader health data dashboard series, which aims to make health-related information more accessible and actionable\footnote{DHHS health data dashboards: \href{https://dhhs.utah.gov/health-dashboards/}{dhhs.utah.gov/health-dashboards/}}.
Within this broader dashboard ecosystem, the wastewater dashboard communicates SARS-CoV-2 surveillance data collected from wastewater sampling sites across the state. The dashboard serves multiple audiences. For DHHS staff, including epidemiologists, it supports monitoring trends, identifying potential outbreaks, informing public health decisions, and reporting surveillance data to the public. For members of the public, it provides access to wastewater-based disease trends in their communities.
The redesign was a collaboration between the University of Utah and DHHS. The University of Utah team included visualization and accessibility researchers and software engineers. We first reviewed the original dashboard for accessibility and usability barriers, developed redesign ideas, and discussed them with DHHS. Based on DHHS feedback and priorities, we refined and implemented the design iteratively. The project lasted roughly one year, with monthly meetings between the University of Utah and DHHS teams. At the end of the project, we conducted a final evaluation interview with DHHS and also collected feedback from an external blind researcher.

\section{Original Dashboard}

\begin{figure*}[t]
    \centering
    \includegraphics[width=1\linewidth, alt={Annotated screenshots compare the original and redesigned Utah Wastewater Surveillance System dashboard on desktop and mobile. The desktop screenshots show the dashboard after a sampling site has been selected. Labels identify key components: (A) statewide charts, which were hidden in a tab in the original version; (B) the summary table, which was collapsed in the original version; (C) the interactive map; (D) the date-range filter for the charts; and (E) site-specific charts. The mobile screenshots show the original map-based landing page and summary table, followed by the redesigned mobile landing page, which replaces the table with a card-based summary view for sampling sites.}]{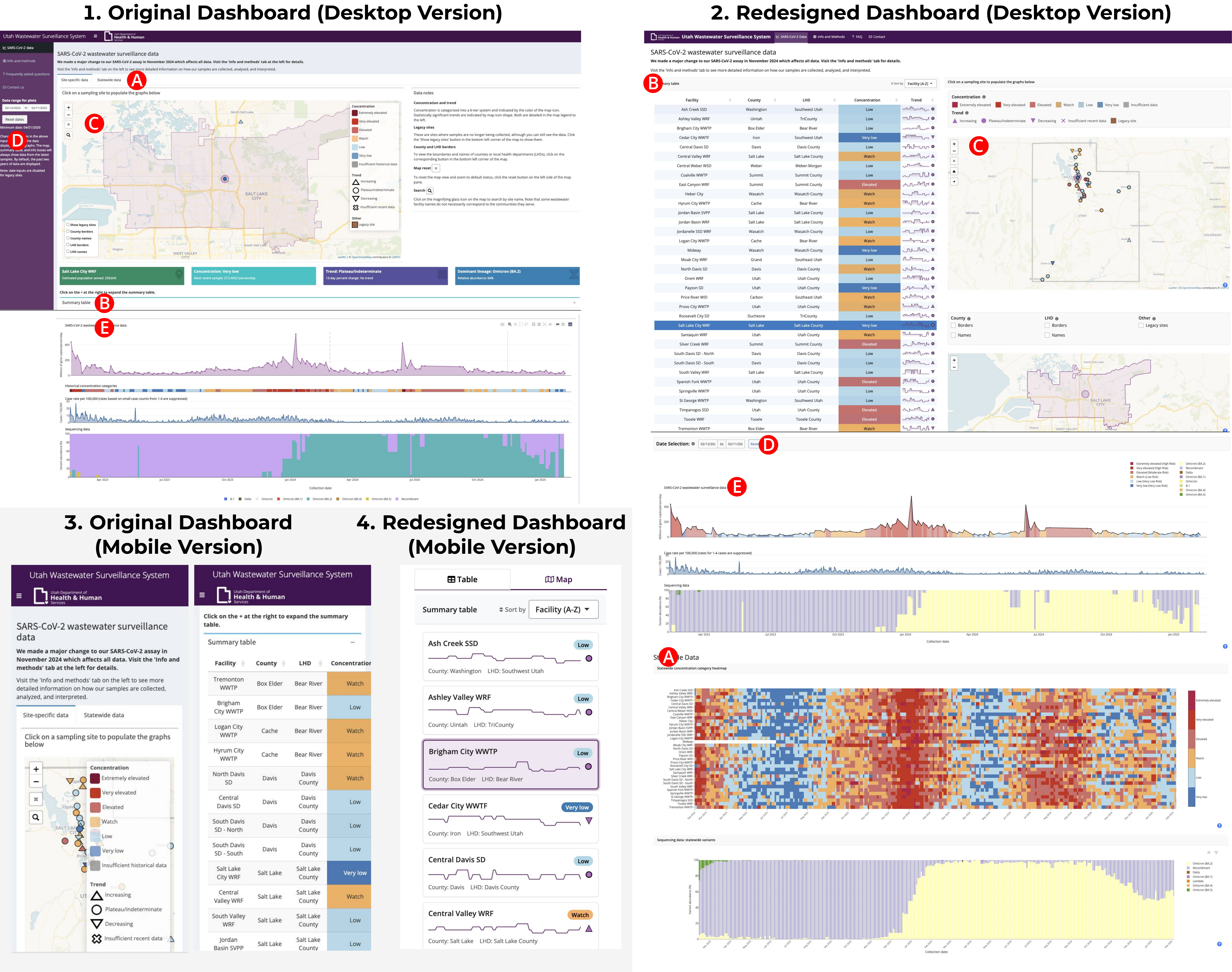}
    \caption{Screenshots of the original and redesigned dashboards on desktop (panels 1 and 2) and mobile (panels 3 and 4). The desktop views show the dashboard after a site has been selected. Labeled features include: (A) statewide charts, which were hidden in a tab in the original version; (B) the summary table, which was collapsed in the original version; (C) the interactive map; (D) the date-range filter for the charts; and (E) site-specific charts. The mobile views show the original map-based landing page and summary table, and the redesigned card-based summary landing page.}
    \label{fig:annotated-screenshots}
\end{figure*}

The original dashboard was a Shiny app~\cite{shiny}, with Plotly~\cite{plotly} for visualizations and Leaflet~\cite{leaflet} for the map. Screenshots of the original dashboard are shown in \autoref{fig:annotated-screenshots}. 
The main view was divided into two tabs: Site-specific data and Statewide data. The Site-specific data tab was the landing view. It displayed an interactive map of 35 Utah wastewater sampling sites, where each site was represented by a colored shape encoding both concentration category and trend status. A data-notes panel next to the map explained the encodings, legacy sites, and available interactions. Below the map, a collapsed summary table listed each site’s current concentration category and trend status.

Selecting a site from the map displayed site-specific information and charts below, including summary cards for population served, latest concentration, current trend, and dominant lineage. It also showed four time-based visualizations: SARS-CoV-2 wastewater concentration, historical concentration levels, reported case rates, and variant abundance over time. Users could adjust the date range in the sidebar to filter these visualizations.

The Statewide data tab provided aggregate views across all sites, including a concentration-category heatmap and a statewide stacked bar chart of variant abundance over time.

\section{Website Redesign}

In this section, we describe the accessibility and usability barriers identified in the original dashboard, the design solutions developed to address them, the iteration process, and DHHS feedback on the final design. \rev{Overall, the redesign prioritized making the information hierarchy clear and easy to access, keeping interactions simple and coordinated, and maintaining consistency across the interface.}

\subsection{Improving the Information Architecture}

\textit{The original dashboard used an inadequately labeled interactive map as the main entry point for site-specific data (see \autoref{fig:annotated-screenshots}, panel 1C).} This structure created an accessibility barrier because selecting a site from a map is difficult for screen-reader users. In addition, the color encodings on the map lacked textual labels, making the encoded information harder to interpret for users with color vision deficiencies. It also created a usability barrier for other users because the map encoded both concentration category and trend status on one symbol, requiring users to interpret the legend before they could identify site-level conditions. Although the original dashboard included a summary table, it was collapsed by default and therefore was not very visible.

We redesigned the landing view around a data-table-first structure, in which the summary table becomes the primary entry point, and the map is retained as a secondary geographic overview next to the table (see \autoref{fig:annotated-screenshots}, panel 2). This structure makes the information more accessible to screen-reader users because the data are presented in a structured textual format rather than only through map-based interaction. Because the dashboard includes only 35 sampling sites, presenting the full table directly does not overload the page. The redesign also benefits sighted users because the table provides a direct textual overview of concentration levels and trend status, which is easier to scan than the visual encodings on the map alone. An additional benefit of emphasizing the table is that it doesn't suffer from the issues maps often have when densely populated areas are overrepresented in the data, and sparsely populated ones have very few data points, an issue especially pronounced in Utah, where the population is highly concentrated. 

DHHS described the consolidated landing view as clean and logically organized, and noted that eliminating secondary tabs could reduce user confusion. They also suggested that future versions could add a pointer to the charts below to make them easier to find on a long page.

\subsection{Linking the Map and Summary Table}

\textit{The original dashboard allowed users to select a site only through the map, while the summary table displayed the same site information but did not support selection.} This interaction made it hard for screen-reader users to make the selection. It was especially problematic after the redesign made the table the primary entry point, because users would reasonably expect to select a site by clicking a row in the table. The limitation came partly from the default behavior of the Shiny components used in the dashboard, which did not directly support bidirectional linking between the table and map.

To address this issue, we created an R package, \texttt{linkeR}, to support bidirectional links between interactive components in Shiny applications. In the redesigned dashboard, users can select a site from either the table or the map (see \autoref{fig:annotated-screenshots}, panels 2B and 2C). The selected site is highlighted across the linked views, making the relationship between the table, map, and detailed plots explicit. This linking supports accessibility because users are no longer required to interact with the map to access site-specific data. It also improves usability because the table and map become coordinated views rather than independent components. 

DHHS described the linked table-map interaction as a ``great addition'' which matches users’ expectation that both views of the same sites should be selectable. They noted that this was a long-desired capability that had been difficult to implement with their existing resources, and they would also use it in their other dashboards.

\subsection{Redesigning the Summary Table}

\textit{The original summary table used encodings that were inconsistent with the map.} For example, trend status was shown with arrows in the table but with shape symbols on the map. This required users to learn separate representations for the same data.
In the redesign, the summary table uses encodings consistent with the map. 
% Concentration categories are shown with color and text labels, and trend status uses the same shape-based encodings as the map. 
We also added sparklines to preview recent trends at each site (see \autoref{fig:annotated-screenshots}, panel 2B). 
When users select a row, the row is marked as selected for screen readers. Its accessible label includes the facility name, county, local health department, concentration category, and trend status.

DHHS responded favorably to the sparklines and described them as a feature they had long wanted to implement. They also suggested clearer tooltips describing both the data shown in the sparklines (categorical concentration levels) and the displayed time range.

\subsection{Redesigning the Map}

\textit{The original map used an unnecessary animated transition when zooming to a selected sampling site.} When users selected a site, the map zoomed to the site, and the site's boundary appeared after the transition. The transition animation was very confusing and did not add the necessary information for interpreting the data. It also created an accessibility concern because the animated interaction made the selected state of the map harder to follow. 

We initially proposed removing the animation and showing the selected site boundary directly. However, during discussions with DHHS, we learned that this was difficult to implement in the existing Leaflet.js-based map. We therefore revised the design: selecting a site highlights it on the main map, and a smaller secondary map below shows the selected site and its boundary directly (see \autoref{fig:annotated-screenshots}, panel 2C). This avoids the confusing transition and supports both the statewide overview and the selected-site boundary view.

\textit{The original map allowed users to zoom out to scales that were not meaningful for a Utah wastewater surveillance dashboard.} Users could zoom out to a world-scale view, which did not support interpretation of the 35 Utah sampling sites. We therefore limited zooming to two useful scales: a Utah statewide view and a Wasatch region view, where sampling sites are more densely located. We also added buttons for these views and reorganized the map legends and controls to improve clarity.

DHHS viewed the revised map as simpler while retaining the current map’s effectiveness. They also emphasized that preserving site boundary information is important for residents who want to determine whether they are included in a sampling area.

\subsection{Redesigning the Visualizations}

\textit{The original site-specific charts used the default Plotly interaction toolbar, which included interaction icons deemed unnecessary for the dashboard's intended use.} The toolbar included twelve interaction icons, such as zoom, pan, box select, lasso select, zoom in, zoom out, autoscale, and reset axes (see \autoref{fig:annotated-screenshots}, panel 1E, top right). Many of these interactions were unnecessary because the dashboard already provides a date-range filter, and the multiple selection, zoom, and navigation methods were unlikely to be clear to general public users. The full toolbar also increased visual complexity and was not accessible to screen-reader users.

We initially proposed removing unnecessary Plotly controls and only using the date-range filter for range selection. However, after discussion with DHHS, we learned that they were familiar with box selection and wanted to keep it. The final design therefore retains reset axes and box selection while removing the other toolbar icons. Box selection remains available through direct dragging on the chart rather than through a visible toolbar icon.

\textit{The original dashboard separated two closely related site-specific charts: the concentration levels area chart and the categorical timeline derived from those values.} This separation required users to compare related quantitative and categorical information across two charts. In the redesign, we explored ways to combine these data in a single visualization, including background category colors, line colors, threshold bands, and threshold lines. 

Based on feedback from a team member with color-vision deficiency, we avoided encoding categories primarily through line color. DHHS initially preferred the background-color design (see \autoref{fig:annotated-screenshots}, panel 2E), but in the final review they reconsidered and noted that color changes occurring at data points might not be intuitive and that background shading might draw attention away from the data. They leaned toward an alternative that uses threshold lines.

\subsection{Alt Text and Downloadable Data}

\textit{The original dashboard did not provide alt text, nor did it offer an option to download the underlying data.} These omissions limited access for screen-reader users and for users who needed to inspect or analyze the data outside the dashboard.

In the redesign, we added alt text for the map and charts (see the blue question-mark icons in charts in \autoref{fig:annotated-screenshots}, panel 2). We made these descriptions visible to all users rather than available only to screen-reader users. Visible alt text might also support users who are less familiar with visualizations in interpreting these charts. We initially proposed using a large language model (LLM) to generate chart alt text, but DHHS did not permit LLM-generated text in the dashboard. We therefore wrote the alt text manually for the final redesign. As a future direction, we propose that template-based dynamic descriptions could provide dynamic alt text without relying on LLM-generated content~\cite{McNutt:2025:Accessible}. We also added data table download functions for the charts.

DHHS emphasized that alt text is a very important part of its accessibility work. They also supported downloadable data as a useful option, but they noted that this feature requires internal approval before deployment.

\section{Mobile Adaptation}

The original dashboard was primarily designed for desktop use, providing little support for mobile devices. Because many users may access this website from their phones, we treated supporting mobile access as part of accessibility, in particular for users who don't have access to a desktop computer. We therefore proposed mobile adaptations to extend the desktop website redesign to smaller touch screens (see \autoref{fig:annotated-screenshots}, panels 3 and 4).

\subsection{Mobile Layout}

\textit{The desktop redesign places the summary table and map side by side, which does not translate well to mobile screens.} 
We therefore proposed stacking the summary view and the map rather than placing them side by side. A tabbed control allows users to switch between the summary view and the map. This design keeps both views available near the top of the dashboard. It also preserves the design rationale of the desktop version: the summary data remains the primary entry point, and the map remains available as a secondary geographic overview. 
For the mobile views, we also omitted the secondary site-boundary map to save space.
This change improves mobile usability by reducing layout complexity and making each view large enough for touch-based interaction. DHHS responded positively to the tabbed mobile layout. They noted that it looked clear and that the summary table worked well in this format.

\subsection{Card View of Summary Table}

\textit{The desktop summary table is difficult to use on mobile devices because it contains too many columns and too much text for a small screen.} To improve readability and touch interaction, we redesigned the mobile summary table as a set of cards. Each card represents a sampling site, with the sparkline placed at the center and related site information organized around it. Compared with table rows, cards provide larger touch targets, better support visual prioritization, and allow secondary attributes to be de-emphasized without removing them. DHHS responded favorably to the card-based summary view, and described it as clean and well-suited to the mobile format.

\subsection{Mobile Visualization Design}

\textit{The statewide time-based visualizations are difficult to adapt to mobile screens because the desktop heatmaps and sequencing charts are wide.} 
We therefore proposed mobile-specific adaptations for these visualizations.
For the statewide heatmap, we considered two options. The first splits the large heatmap into smaller county-level heatmaps, moves the site legend below the chart to save horizontal space, and adds a filter to support navigation. This improves mobile readability but differs from the desktop representation. The second option stays closer to the desktop design by limiting the displayed time range, such as to two months, and moving the legend below the chart. However, it still leaves limited space for site labels on narrow screens. We presented both designs to DHHS, who preferred the first option and were comfortable with mobile visualizations differing from the desktop version as long as the mobile view remained clear and easy to navigate.

The sequencing visualization raised a related mobile design problem because its legend uses seven colors to represent different viral lineages. A full, always-visible legend supports interpretation and comparison, but it consumes valuable screen space on a phone. 
We considered two legend options: an on-demand legend and a minimal legend showing only the categories present in the current chart. DHHS supported the minimal legend design, but this behavior is not directly supported by the Plotly components used in the existing dashboard. We therefore considered developing an additional wrapper around Plotly to preserve DHHS’s existing visualization library, improve maintainability, and provide greater control over mobile adaptations.

\subsection{Evaluation}

\paragraph{Evaluation with DHHS collaborators}
We conducted a semi-structured interview with our DHHS collaborators to collect feedback on the redesigned dashboard. Overall, DHHS viewed the redesign as useful and aligned with their goals for improving the dashboard. They stated that the collaboration helped refine the presentation of wastewater data, implement technical capabilities for which they lacked capacity, and support their ADA accessibility efforts. We also report their feedback for each specific redesign component in the corresponding subsections above. Although DHHS suggested several small improvements, they characterized the remaining issues as minor refinements rather than major barriers. Regarding deployment, DHHS indicated that deployment is on their to-do list and is limited by staff time rather than technical barriers.

\paragraph{Evaluation with a Blind Researcher}
We conducted a brief review with an external blind researcher, who compared the original and redesigned dashboards and provided written feedback by email. The feedback highlighted that, while placing the summary table upfront was a useful direction, several barriers still limited the dashboard’s practical usefulness for screen-reader users, including chart descriptions being too general (describing only the encoding but not the data), interaction issues, and element labeling problems. 
\rev{We implemented changes to address the labeling problems and many of the interaction issues identified in the review. Evaluating these changes with screen-reader users, as well as addressing more structural interaction issues and providing dynamic alt text for interactive charts (as discussed previously), remain directions for future work.}

\section{Discussion}

This case study highlights the difficulty of implementing ambitious accessibility solutions in a constrained public-sector environment \rev{and within existing government workflows}. The redesign had to work within the existing toolchain, DHHS approval processes, limited staff time, data-access constraints, and restrictions on LLM-generated content. \rev{Our experience suggests that, in this context,} simple technical solutions tailored to the existing environment, such as the \texttt{linkeR} package \cite{linkeR}, can provide substantial practical value. 

\rev{Our redesign was informed by WCAG and aimed to address accessibility barriers relevant to DHHS's ADA obligations, but the final dashboard was not formally evaluated for WCAG conformance. Formally evaluating the redesign against WCAG therefore remains future work. The blind researcher’s feedback suggests that guideline-informed accessibility improvements, while important, may still leave barriers that affect the practical usefulness of a dashboard for blind and low-vision users. This gap is important for designers of accessible dashboards to consider, particularly for dashboards that combine multiple visualizations and interactive controls. Our current evaluation, based on feedback from a single blind researcher, represents only an initial step. Future work includes broader evaluations with disabled users, including task-based evaluations. It is also important to study how to address remaining access needs, such as better support for interactive visualizations, dynamic alt text, and multilingual access (e.g., Spanish-language support).}

\acknowledgments{%
This work was supported by the Centers for Disease Control and Prevention's Center for Forecasting and Outbreak Analytics Cooperative Agreement CDC-RFA-FT-23-0069. We acknowledge the feedback and collaborative efforts of Nathan LaCross, Kerry Regan, Mary Jewell, Bree Barbeau, and Abigail Collingwood from the Utah DHHS. We thank Daniel Hajas for his valuable feedback on the redesign and Nate Lanza for his help in developing several dashboard features.
}

\bibliographystyle{abbrv-doi-hyperref-narrow}
\bibliography{abbreviations,references}

\end{document}